\documentclass[a4paper]{jpconf}
\usepackage{graphicx}
\usepackage[percent]{overpic}
\usepackage{cite}

\begin{document}
\title{J/$\psi$ and $\psi(2S)$ measurement in $p$+$p$ collisions at $\sqrt{s} =$ 200 and 500 GeV in the STAR experiment}

\author{Barbara Trzeciak$^1$ for the STAR Collaboration}

\address{$^1$Faculty of Nuclear Sciences and Physical Engineering, Czech Technical University in Prague, Brehova 7, 115 19 Praha 1, Czech Republic}

\ead{trzecbar@fjfi.cvut.cz}

\begin{abstract}
In this paper, results on the J/$\psi$ cross section and polarization measured via the dielectron decay channel at mid-rapidity in $p+p$ collisions at $\sqrt{s}$ = 200 and 500 GeV in the STAR experiment are discussed. The first measurement of $\psi(2S)$ to J/$\psi$ ratio at $\sqrt{s}$ = 500 GeV is also reported. 
\end{abstract}

\section{Introduction}

J/$\psi$ and $\psi(2S)$ are bound states of charm ($c$) and anti-charm ($\overline{c}$) quarks. Charmonium physical states have to be colorless, however they can be formed via a color-singlet (CS) or color-octet (CO) intermediate $c\overline{c}$ state. One of the first models of the charmonium production, the Color Singlet Model (CSM) \cite{Braaten:1996pv}, assumed that J/$\psi$ is created through the color-singlet state only. This early prediction failed to describe the measured charmonium cross section which has led to the development of new models. For example, Non-Relativistic QCD (NRQCD) \cite{Braaten:1996pv} calculations were proposed in which a $c\overline{c}$ color-octet intermediate states, in addition to a color-singlet states, can bind to form charmonia.

However, the charmonium production mechanism in elementary particle collisions is not yet exactly known.
For many years measurements of the J/$\psi$ cross section have been used to test different J/$\psi$ production models. While many models can describe relatively well the experimental data on the J/$\psi$ cross section in $p+p$ collisions \cite{Abelev:2009qaa, Adamczyk:2012ey, Adare:2011vq, PhysRevLett.79.572, Acosta:2004yw, Aad:2011sp, Khachatryan:2010yr, Aaij:2011jh}, they have different predictions for the J/$\psi$ polarization. Therefore, measurements of the J/$\psi$ polarization may allow to discriminate among different models and provide new insight into the J/$\psi$ production mechanism.

\section{Charmonium measurements in STAR}

In STAR, charmonia have been measured so far via the dielectron decay channel. The STAR detector~\cite{Ackermann2003624} is a multi-purpose detector that has large acceptance at mid-rapidity, $\vert \eta \vert <$ 1 with a full azimuthal coverage. Electrons can be identified using the Time Projection Chamber (TPC)~\cite{Anderson:2003ur} through ionization energy loss ($dE/dx$) measurement. The Time Of Flight (TOF) detector~\cite{Llope:2012ti} greatly enhances the electron identification capability at low momenta where the $dE/dx$ bands for electrons and hadrons cross each other. At high $p_{T}$, electron identification can be improved by the Barrel Electromagnetic Calorimeter (BEMC)~\cite{Beddo:2002zx} which measures electron energy and shower shape. The BEMC is also used to trigger on high-$p_{T}$ electrons (HT trigger). Minimum Bias (MB) events are triggered by the Vertex Position Detectors (VPD)~\cite{Llope:2003ti}.

\section{J/$\psi$ measurements in $p$+$p$ at $\sqrt{s} =$ 200 GeV}

STAR has measured inclusive J/$\psi$ $p_{T}$ spectra and polarization in $p+p$ collisions at $\sqrt{s} =$ 200 GeV via the dielectron decay channel ($B_{ee} =$ 5.9\%) at mid-rapidity ($\vert y \vert < $ 1). These results are compared to different model predictions to understand J/$\psi$ production mechanism in elementary collisions.

\begin{figure}[ht]
		\includegraphics[width=0.42\linewidth]{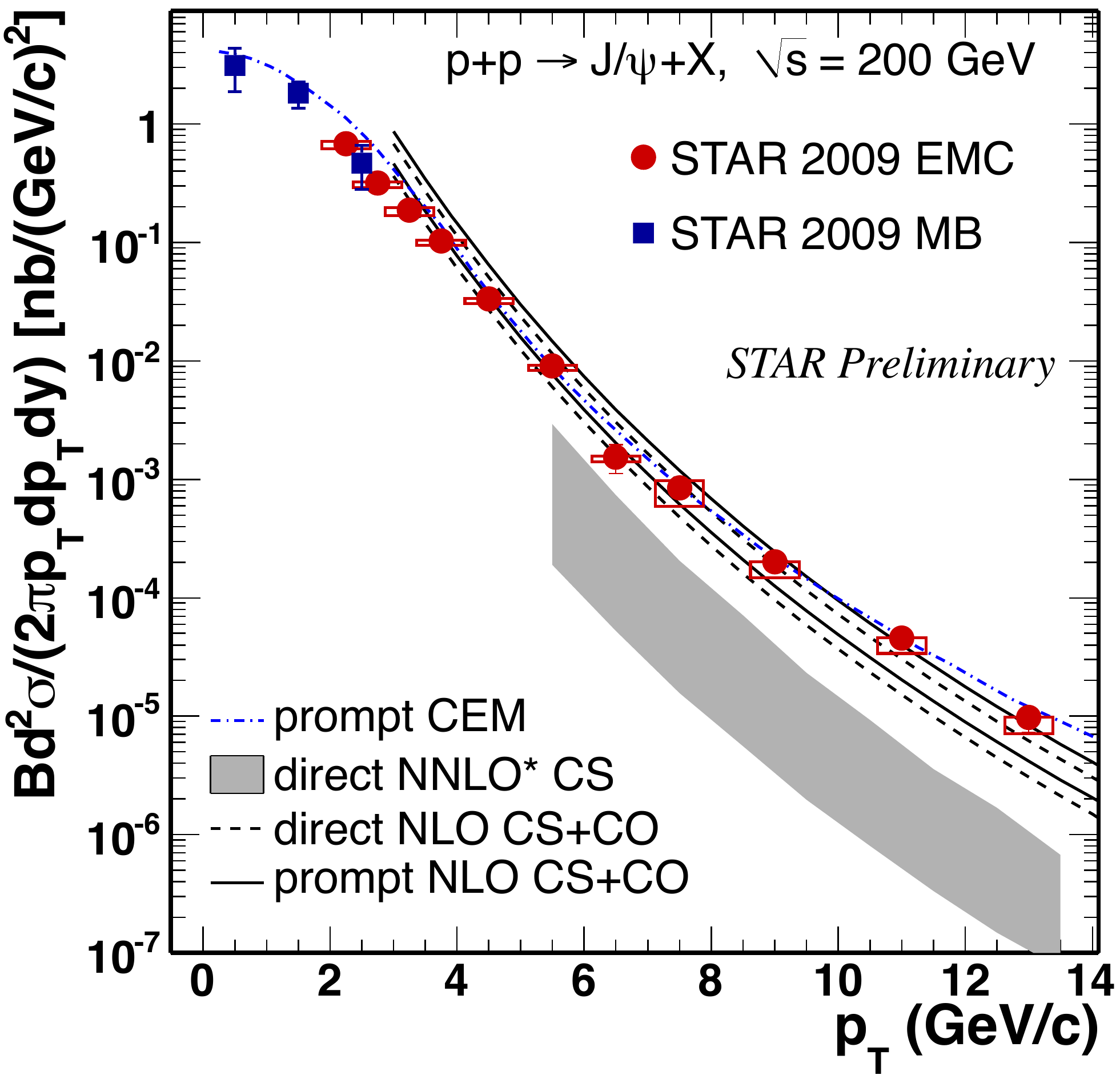}
		\includegraphics[width=0.58\textwidth]{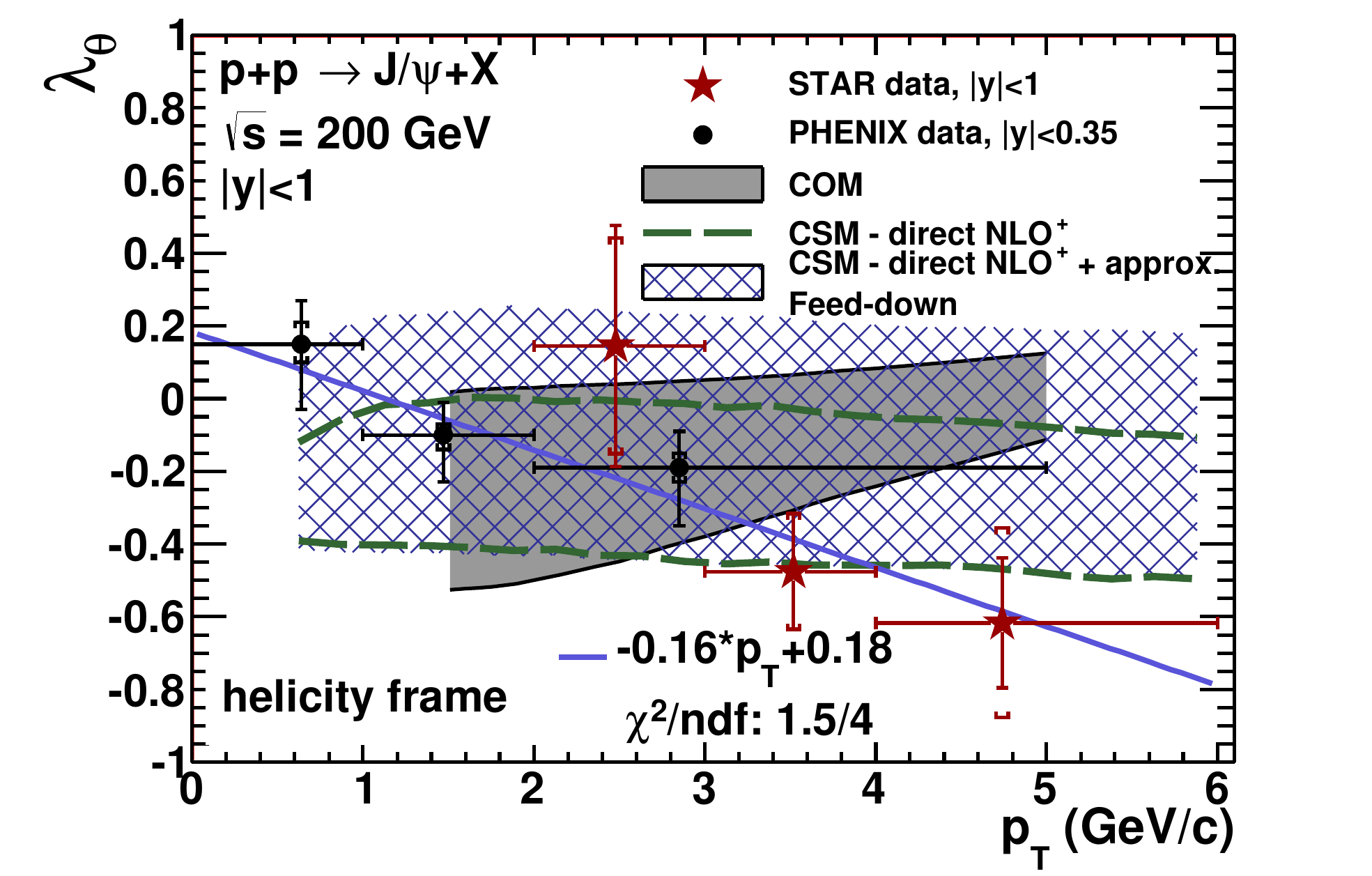}
		\caption{Left: J/$\psi$ invariant cross section vs $p_{T}$ in $p$+$p$ collisions at $\sqrt{s} =$ 200 GeV at mid-rapidity at low~\cite{Kosarzewski:2012zz} and high $p_{T}$~\cite{Adamczyk:2012ey} shown as blue squares and red circles, respectively, compared to different model predictions~\cite{Frawley:2008kk,Ma:2010jj,Artoisenet:2008fc}. Right: Polarization parameter $\lambda_{\theta}$ vs J/$\psi$ $p_{T}$ for $|y| <$ 1~\cite{Adamczyk:2013vjy} compared to the PHENIX measurement~\cite{PhysRevD.82.012001} and two model predictions~\cite{PhysRevD.81.014020,Lansberg:2010vq}.}
		\label{fig:Jpsi200}
\end{figure}

Left panel of Fig.~\ref{fig:Jpsi200} shows STAR low and high-$p_{T}$ measurements of J/$\psi$ $p_{T}$ spectra~\cite{Kosarzewski:2012zz, Adamczyk:2012ey} compared to model predictions. The Color Evaporation Model (CEM)~\cite{Frawley:2008kk} for prompt J/$\psi$ can describe the $p_{T}$ spectrum reasonably well, except the region around $p_{T}\approx$ 3 GeV/$c$ where it over-predicts the data. NLO NRQCD calculations with color-singlet and color-octet transitions \cite{Ma:2010jj} for prompt J/$\psi$ match the data for $p_{T} > 4$ GeV/c. NNLO* CS model \cite{Artoisenet:2008fc} for direct J/$\psi$ production under-predicts the STAR data, but the prediction does not include contributions from $\psi(2S)$, $\chi_{C}$ and $B$-meson decays to J/$\psi$.

In $p$+$p$ collisions at $\sqrt{s} =$ 200 GeV STAR has also measured J/$\psi$ polarization parameter $\lambda_{\theta}$ in the helicity frame at mid-rapidity and 2 $ < p_{T} <$ 6 GeV/$c$~\cite{Adamczyk:2013vjy}. 
$J/\psi$ polarization is analyzed via the angular distribution of the decay electrons that is described by:
$\frac{d^{2}N}{d(\cos\theta)d\phi} \propto 1+\lambda_\theta \cos^2\theta + \\ \lambda_\phi \sin^2\theta \cos2\phi + \lambda_{\theta\phi}\sin2\theta \cos\phi$, where $\theta$ and $\phi$ are polar and azimuthal angles, respectively; $\lambda_\theta$, $\lambda_\phi$ and $\lambda_{\theta\phi}$ are the angular decay coefficients. 
The $p_{T}$ dependence of $\lambda_{\theta}$ is shown on the right panel of Fig.~\ref{fig:Jpsi200} with low-$p_{T}$ PHENIX results~\cite{PhysRevD.82.012001} and compared to NRQCD calculations~\cite{PhysRevD.81.014020} and the NLO$^{+}$ CSM prediction~\cite{Lansberg:2010vq}. A trend observed in the RHIC data is towards longitudinal polarization as $p_{T}$ increases and, within experimental and theoretical uncertainties, the result is consistent with the NLO$^{+}$ CSM model.

\begin{figure}[ht]
		\includegraphics[width=0.5\linewidth]{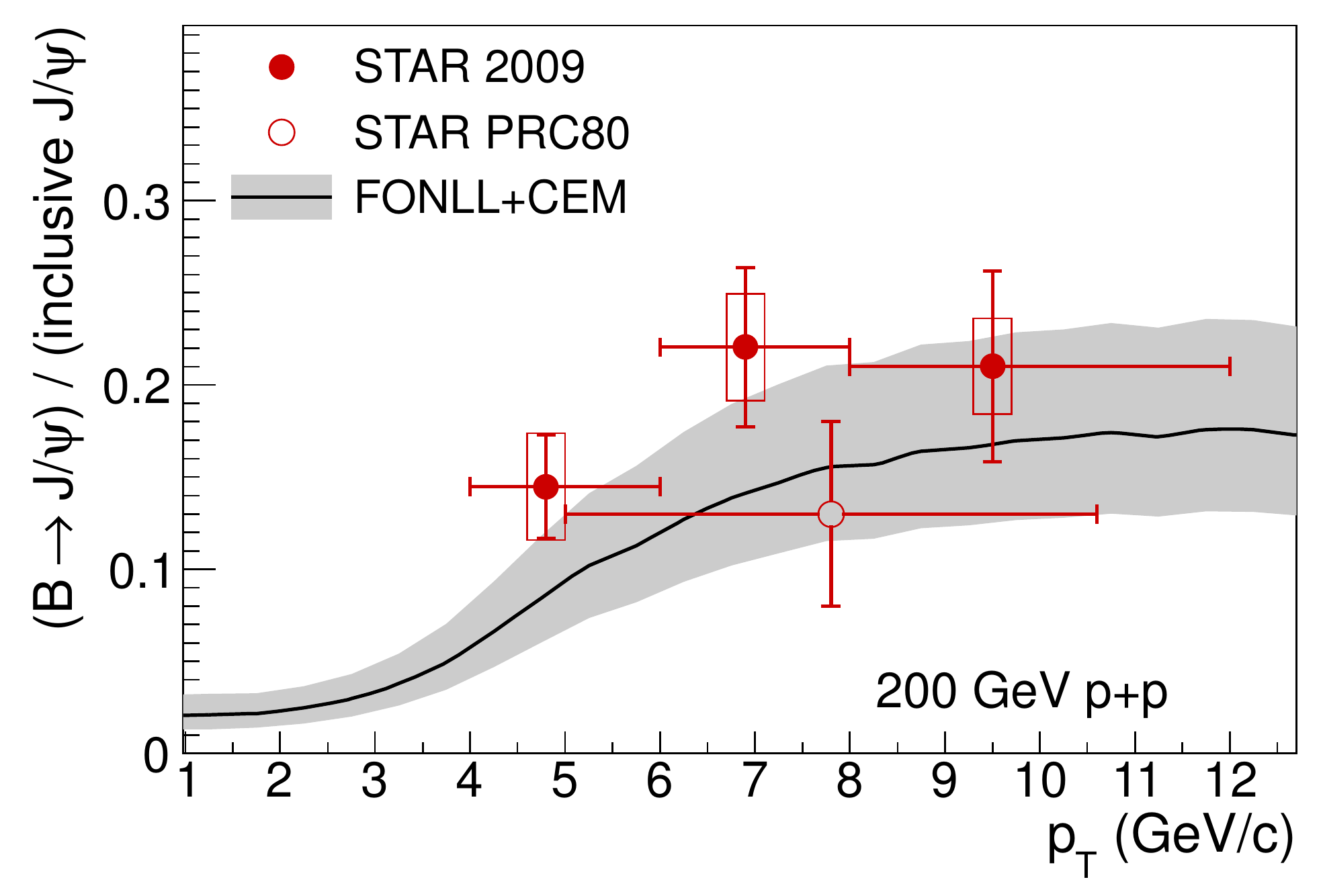}
		\includegraphics[width=0.5\linewidth]{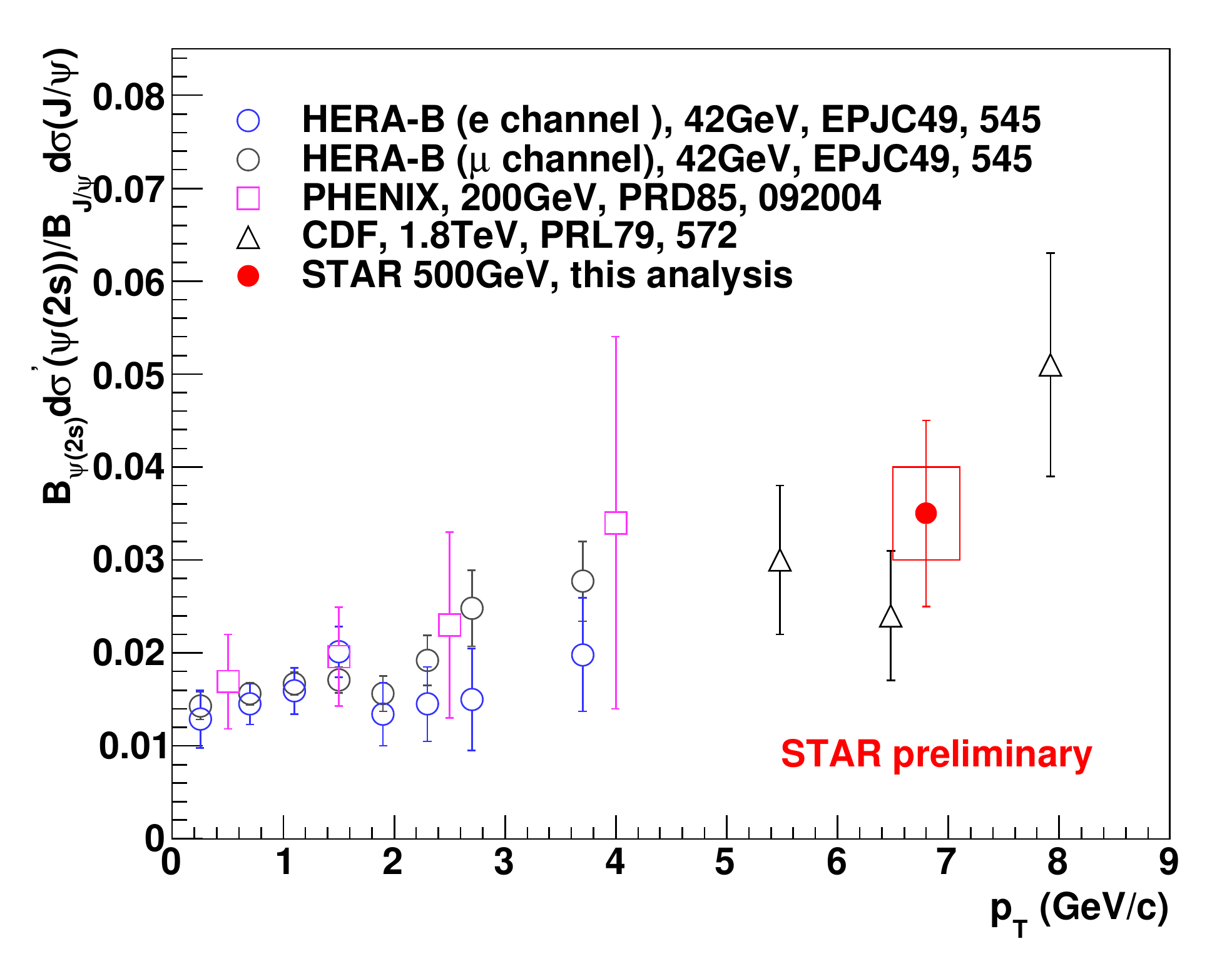}
		\caption{Left: relative contribution from B-meson decays to inclusive J/$\psi$ production in $p$+$p$ at $\sqrt{s} =$ 200 GeV~\cite{Adamczyk:2012ey} compared to FONLL+CEM calculations~\cite{Bedjidian:2004gd,Cacciari:2005rk}. Right: ratio of $\psi(2S)$ to J/$\psi$ in $p+p$ collisions at $\sqrt{s} =$ 500 GeV from STAR (red circle) compared to results from other experiments at different energies.}
		\label{fig:feedDown}
\end{figure}

The inclusive J/$\psi$ production is a combination of prompt and non-prompt J/$\psi$. The prompt J/$\psi$ production consists of the direct one ($\sim$60\%) and feed-down from excited states $\psi(2S)$($\sim$10\%) and $\chi_{C}$($\sim$30\%), while non-prompt J/$\psi$ originate from B-hadron decays. STAR has estimated the contribution from B-meson decays using a measurement of azimuthal angular correlation between high-$p_{T}$ J/$\psi$ and charged hadrons~\cite{Abelev:2009qaa,Adamczyk:2012ey}. The relative contribution of B-hadron decays to inclusive J/$\psi$ yield is strongly $p_{T}$ dependent and it is 10-25\% for 4 $< p_{T} <$ 12 GeV/$c$, as it is shown on the left panel of Fig.~\ref{fig:feedDown}. The measurement is consistent with the FONLL+CEM prediction~\cite{Bedjidian:2004gd,Cacciari:2005rk}.

\section{J/$\psi$ and $\psi(2S)$ measurements in $p$+$p$ at $\sqrt{s} =$ 500 GeV}

\begin{figure}[ht]
		\center
		\includegraphics[width=0.45\linewidth]{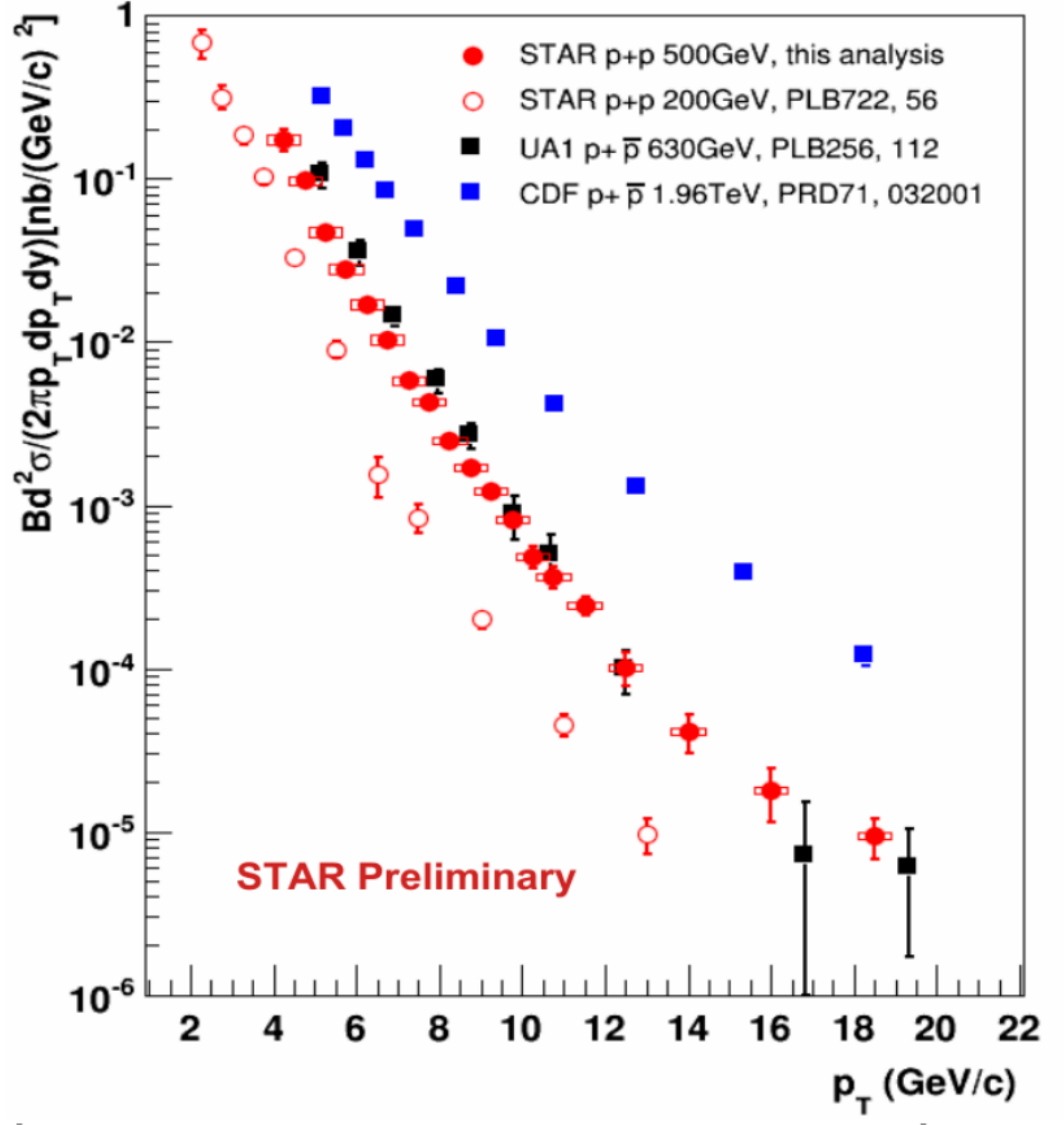}
		\hspace{20pt}
		\includegraphics[width=0.45\linewidth]{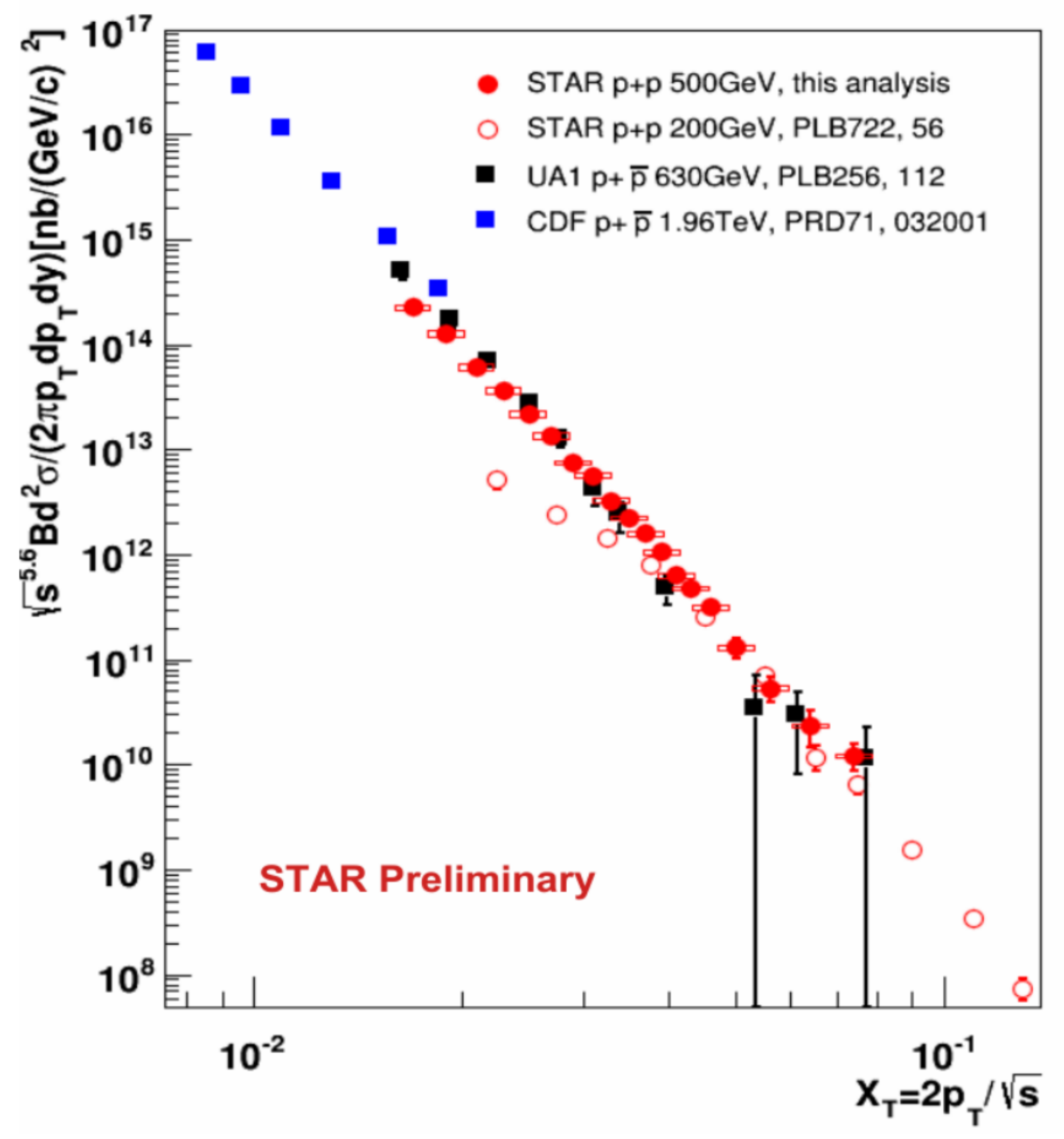}
		\caption{J/$\psi$ invariant cross section vs $p_{T}$, left panel, and invariant cross section multiplied by $\sqrt{s}^{5.6}$ vs $x_{T}$, right panel, in $p$+$p$ collisions at $\sqrt{s} =$ 500 GeV at mid-rapidity shown as full circles compared to measurements at different energies.}
		\label{fig:Jpsi500}
\end{figure}

In order to further test the charmonium production mechanism and constrain the feed-down contribution from the excited states to the inclusive J/$\psi$ production, the J/$\psi$ and $\psi(2S)$ signals were extracted in $p+p$ collisions at $\sqrt{s} =$ 500 GeV at mid-rapidity. The J/$\psi$ $p_{T}$ spectrum is shown on the left panel of Fig.~\ref{fig:Jpsi500}. The STAR results at $\sqrt{s} =$ 500 GeV (full circles) are compared to those at $\sqrt{s} =$ 200 GeV (open circles) and with measurements of other experiments in $p$+$\bar{p}$ collisions at different energies. The STAR measurements cover $p_{T}$ range of 4 - 20 GeV/$c$ with a good precision. 
It was also observed that J/$\psi$ cross section follows the $x_{T}$ scaling: $\frac{d^{2}\sigma}{2 \pi p_{T} dp_{T} dy} = g(x_{T})/(\sqrt{s})^{n}$, where $x_{T} = 2p_{T}/\sqrt{s}$, with $n =$ 5.6 $\pm$ 0.2 at mid-rapidity and $p_{T} >$ 5 GeV/$c$ for a wide range of colliding energies~\cite{Abelev:2009qaa}. At $\sqrt{s} =$ 500 GeV the same $x_{T}$ scaling of high-$p_{T}$ J/$\psi$ production is seen, as shown on the right panel of Fig.~\ref{fig:Jpsi500}.

Right panel of Fig.~\ref{fig:feedDown} shows $\psi(2S) / J/\psi$ ratio from STAR (red full circle) compared to measurements of other experiments at different colliding energies, in $p+p$ and $p+$A collisions. The STAR data point is consistent with the observed trend, and no collision energy dependence of the $\psi(2S)$ to J$/\psi$ ratio is seen with the current precision.

The statistics available at $\sqrt{s} =$ 500 GeV will allow us to extract the frame invariant polarization parameter, also in different reference frames, providing model independent information about the J/$\psi$ polarization~\cite{Faccioli:2010kd}. It will be possible to measure the azimuthal polarization parameter, $\lambda_{\phi}$, and improve precision of the $\lambda_{\theta}$ measurement. Analysis of J/$\psi$ polarization at $\sqrt{s} =$ 500 GeV is ongoing.

\section{Summary}

In summary, STAR has measured the inclusive J/$\psi$ cross section and polarization in $p$+$p$ collisions at $\sqrt{s} =$ 200 GeV as a function of $p_{T}$. The measurements are compared to different model predictions of the J/$\psi$ production. The $p_{T}$ spectrum is described well by the NRQCD calculations while the measured polarization parameter $\lambda_{\theta}$ is consistent with the NLO$^{+}$ CSM prediction. STAR new result for J/$\psi$ at $\sqrt{s} =$ 500 GeV extends $p_{T}$ reach up to 20 GeV/$c$. 
The first measurement of $\psi(2S) / J/\psi$ ratio in $p$+$p$ collisions at $\sqrt{s} =$ 500 GeV is reported and compared with results from other experiments. No collision energy dependence is observed. 

\section*{Acknowledgements}

This publication was supported by the European social fund within the framework of realizing the project ,,Support of inter-sectoral mobility and quality enhancement of research teams at Czech Technical University in Prague'', CZ.1.07/2.3.00/30.0034. 

\section*{References}

\bibliographystyle{iopart-num} 
\bibliography{/home/barbara/Work/bib/Bibliography_all}

\end{document}